\documentclass[12pt]{article}

\def\slash#1{\setbox0=\hbox{$#1$}#1\hskip-\wd0\hbox to\wd0{\hss\sl/\/\hss}}
\usepackage{epsf, cite}
\usepackage{epsfig}                  
\usepackage[all]{xy}                 
\usepackage{amsfonts}
\usepackage{latexsym}
\usepackage{amsmath}
\usepackage{amssymb}

\usepackage{epsf, cite}

\makeatletter
\renewcommand\section{\@startsection {section}{1}{\z@}%
                                   {-3.5ex \@plus -1ex \@minus -.2ex}
                                   {2.3ex \@plus.2ex}%
                                   {\normalfont\large\bfseries}}
\renewcommand\subsection{\@startsection{subsection}{2}{\z@}%
                                     {-3.25ex\@plus -1ex \@minus -.2ex}%
                                     {1.5ex \@plus .2ex}%
                                     {\normalfont\bfseries}}
\makeatother

\let\non\nonumber

\def\lbldef#1#2{\expandafter\gdef\csname #1\endcsname {#2}}

\def\href#1#2{#2}

\def\beq{\begin{equation}}
\def\eeq{\end{equation}}
\def    \bea    {\begin{eqnarray}}
\def    \eea    {\end{eqnarray}}

\newcommand{\Ab}{{\bar{A}}}
\newcommand{\Bb}{{\bar{B}}}
\newcommand{\Cb}{{\bar{C}}}
\newcommand{\Db}{{\bar{D}}}
\newcommand{\Eb}{{\bar{E}}}

\def\ab{{\bar{\alpha}}}
\def\bb{{\bar{\beta}}}

\def\db{{\bar{\delta}}}
\def\dlb{{\bar{\delta}}}

\def\cb{{\bar{\sigma}}}

\def\ar{{\bar{a}}}
\def\br{{\bar{b}}}

\def\dr{{\bar{d}}}
\def\er{{\bar{e}}}

\def\dtp{\frac{d^2p}{(2\pi)^2}}
\renewcommand{\a}{\alpha}
\renewcommand{\b}{\beta} 
\renewcommand{\c}{\sigma}
\def\s{\sigma}
\newcommand{\dl}{\delta}
\newcommand{\e}{\epsilon}
\renewcommand{\t}{\theta}
\def\g{\gamma}
\def\m{\mu}

\renewcommand{\H}{\mathcal{H}}
\def\M{\mathcal G}
\def\L{\mathcal L}
\def\K{\mathcal K}
\def\V{\mathcal V}
\newcommand{\X}{\mathbb{X}}
\renewcommand{\d}{\partial} 
\newcommand{\G}[3]{\Gamma^{#1}_{\ #2 #3}}
\newcommand{\Vt}[2]{\V^{#1}_{\ #2}}
\newcommand{\p}{\prime}
\def\Dh{\hat{D}}
\providecommand{\openone}{\leavevmode\hbox{\small1\kern-3.8pt\normalsize1}}
\def\xt{\tilde{X}}
\newcommand{\uc}{{\underline{\c}}}

\begin{document}
\pagestyle{plain}
\begin{titlepage}

\begin{center}

\hfill{QMUL-PH-2007-15} \\

\vskip 1cm

{{\Large \bf Background Field Equations for the Duality Symmetric String}} \\

\vskip 1.25cm {David S. Berman\footnote{email: D.S.Berman@qmul.ac.uk},
  Neil B. Copland\footnote{email: N.B.Copland@qmul.ac.uk} } and
Daniel C. Thompson\footnote{email: D.C.Thompson@qmul.ac.uk}
\\
{\vskip 0.2cm
Queen Mary College, University of London,\\
Department of Physics,\\
Mile End Road,\\
London, E1 4NS, England\\
}

\end{center}
\vskip 1 cm

\begin{abstract}
\baselineskip=18pt\

This paper describes the background field equations for strings in
T-duality symmetric formalisms in which the dimension of target space
is doubled and the sigma model supplemented with constraints. 
These are calculated by demanding the vanishing of 
the beta-functional of the sigma model couplings in the doubled target
space. We demonstrate the equivalence with the background field equations of the standard string sigma model.

\end{abstract}

\end{titlepage}

\pagestyle{plain}

\baselineskip=19pt

\section{Introduction}

T-duality is one of the cornerstones of string theory. It is an
intrinsically stringy effect which relates small to large manifolds 
exchanging winding with momentum modes. From a space-time 
perspective T-duality is a solution generating
symmetry of the low energy equations of motion but from the world
sheet point of view, T-duality is a
non-perturbative symmetry. Importantly, the presence of
T-duality allows for the construction of non-geometric manifolds
where locally geometric regions are patched together with T-duality
transformations. Such constructions, known as T-folds \cite{tfold1}, may
play a crucial role in moduli stabilisation and certainly are an
important part of any string landscape.

Given the importance of T-duality it is desirable that this symmetry
is made manifest in the string sigma model. 
There have been various attempts in the past to
develop a formalism where T-duality is a symmetry of the action \cite{past}. We will
use the form most recently championed by Hull \cite{hull1} as our starting point. This doubles the
dimension of the target space with the doubled dimensions
being related by T-duality. Additional constraints are then needed to
reduce the degrees of freedom. These take the form of a set of
chirality constraints. The result is that the formalism has manifest
T-duality with a doubled target space. The key difficulty is dealing  with the constraints in the correct way. Demonstrating the classical equivalence of this formalism to the usual sigma model is straightforward, however, showing quantum equivalence is less trivial.   This has been discussed in \cite{bermancopland,emily,Chowdhury:2007ba}. 

In this paper we would like to examine the beta-functional of the
string in this formalism so as to determine the background-field equations for the
doubled space. This project was prompted by various questions. The
most important was to determine if the background-field equations arising from the one-loop beta-functional for
the doubled formalism were the same as for the usual string. A priori
this did not have to be the case.  
In fact, one may imagine that they will be different since the doubled
formalism naturally incorporates the string winding modes which could
in principle correct the usual the beta-function. We know that
world sheet instantons correct T-duality \cite{harvey,tong} and so
since the doubled space contains the naive T-dual one may think of all sorts of
possibilities that could arise for corrections to the doubled geometry. 

We begin by briefly introducing the doubled formalism before 
incorporating the constraints into the action so that we can use 
the background-field method. We then carry out the background-field 
expansion to find the resulting one loop beta-functional
for the background doubled metric. 
Finally the relation to the non-doubled geometry is described. 
This shows that the constraints on the doubled geometry required for conformal invariance are equivalent to
the usual background-field equations for the standard string sigma model.

\section{The doubled formalism}

The doubled formalism\cite{tfold1,hull1,Hull:2006va} is an alternate description of string theory on target spaces that are locally $T^n$ bundles, with fibre coordinates $X^i$, over a base $N$ with coordinates $Y^a$.  The fibre is doubled to be a $T^{2n}$ with $2n$ coordinates $\X^A $. The doubled sigma model then has Lagrangian\footnote{The complete formalism also introduces a 1-form connection for the fibration $\mathcal{A}^A(Y)$ which we set to zero throughout this paper.}
\beq
\mathcal{L} = \frac{1}{4} \H_{AB}(Y)d\X^A\wedge\ast d\X^B  +
\mathcal{L}(Y) + \mathcal{L}_{top}(\X)
\eeq
where $\mathcal{L}(Y)$ is the standard Lagrangian for a string on the
base and $\H(Y)$ is a metric on the fibre\footnote{We adopt the conventions of \cite{Hull:2006va}; the worldsheet signature is $(+,-)$, $\d_\pm = \d_0 \pm \d_1$, $\e_{01}= 1 = \e^{01}$ and for convenience we have dropped an overall factor of $2\pi$.  The factor of $\frac{1}{4}$ in (1) is half the usual normalisation and is required to make contact with the standard sigma model.}.
$\mathcal{L}_{top}$ is a purely topological term described in
\cite{Hull:2006va}. It is vital
to obtaining the equivalence of the doubled to the non-doubled partition functions \cite{bermancopland}
but it will play no role in the beta-functions and so will be dropped from now on. One may choose a frame where the metric $\H$ has an $O(n,n)/O(n)\times
O(n)$ coset form as follows:
\bea\label{H}
\H_{AB}(Y) &=& \left( \begin{array}{cc}
h - bh^{-1}b & bh^{-1}\\
-h^{-1}b & h^{-1}
\end{array}\right)\, .
\eea
$h$ and $b$ are the target space metric and $B$-field on the fibre of
the undoubled space. In this frame $\X^A=(X^i,\xt_i)$ 
with $\{\xt_i\}$ the coordinates on the T-dual torus. 
We must then supplement this doubled sigma model action with a set of constraints so as to obtain the
correct number of degrees of freedom and be equivalent to the usual
non-doubled sigma model. The constraints are given by
\beq
\label{eqConstraint}
d\X^A = L^{AB}\H_{BC}\ast d\X^C ,
\eeq
where the $L$ is an $O(n,n)$ invariant metric that can be used to raise and lower indices on $\H$ which in this basis is given by
\bea  L_{AB} = \left( \begin{array}{cc}
0& \openone\\
\openone & 0
\end{array}\right).
\eea
 To understand this constraint it is helpful to introduce a vielbein to allow a change to a chiral frame (denoted by barred indices) where:
\bea
\H_{\Ab\Bb}(y) = \left( \begin{array}{cc}
\openone &0\\
0 & \openone
\end{array}\right), &  L_{\Ab\Bb} = \left( \begin{array}{cc}
\openone &0\\
0 & -\openone
\end{array}\right).
\eea
In this frame the constraint (\ref{eqConstraint}) is a chirality
constraint ensuring that half the $\X^\Ab$ are chiral Bosons and half
are anti-chiral Bosons.
\par 
As a simple example let us consider the $n=1$ case ie. a circle, with
constant radius $R$. The doubled action on the fibre is
\beq
S_{d}=\frac{1}{4} R^2 \int dX\wedge \ast dX + \frac{1}{4} R^{-2} \int d\tilde X \wedge \ast d\tilde X .
\eeq
As in \cite{bermancopland}, we make the change to basis in which the fields are chiral:
\bea
P= RX+R^{-1}\tilde{X}, &  \d_-P =0\, , \\ Q= RX-R^{-1}\tilde{X}, & \d_+Q =0\, . 
\eea
In this basis the action becomes
\beq
S_{d}= \frac{1}{8}\int dP\wedge \ast dP + \frac{1}{8} \int d  Q \wedge \ast d Q\, .
\eeq
One may then incorporate the constraints into the action using the
method of Pasti, Sorokin and Tonin \cite{Pasti:1996vs}. We define one-forms
\bea
\mathcal{P}=dP-\ast dP ,& \mathcal{Q}=dQ+\ast dQ\, ,
\eea
which vanish on the constraint.  These allow us to incorporate the constraint into the action via the introduction of two auxiliary closed one-forms $u$ and $v$ as follows: 
\bea
S_{PST}= \frac{1}{8}\int dP\wedge \ast dP+ \frac{1}{8}\int dQ \wedge \ast dQ -\frac{1}{8}\int d^2\sigma\left( \frac{(\mathcal{P}_m u^m)^2}{u^2} +\frac{(\mathcal{Q}_m v^m)^2}{v^2}\right).
\eea

The PST action works by essentially introducing a new gauge symmetry,
the PST symmetry, that allows the gauging away of fields that do not
obey the chiral constraints. Thus only the fields obeying the chiral
constraints are physical.

There are now two ways to proceed. One may either gauge fix the
PST-style action immediately 
which will break manifest Lorentz invariance
or try to quantise covariantly and introduce ghosts to deal
with the PST gauge symmetry. In this paper we choose the non-covariant
option and immediately gauge fix to give a Floreanini-Jackiw\cite{FJ}
style action. Picking the auxiliary PST fields ($u$ and $v$) to be
time-like produces two copies of the FJ action (one chiral and one anti-chiral)
\beq\label{FJL}
S = \frac{1}{4}\int d^2\s ( \d_1P\d_{-}P - \d_{1}Q \d_+Q).
\eeq
We re-expand this in the non-chiral basis to give Tseytlin's \cite{Tseytlin1} duality-symmetric formulation of the action
\beq
\label{eqTseyaction}
 S= \frac{1}{2}\int d^2\sigma\left[ -(R\d_1 X)^2 - ( R^{-1}\d_1 \tilde{X})^2 + 2\partial_0X\partial_1\tilde{X}\right] .
\eeq
Notice that the unusual normalisation of (1) was exactly what was needed for this form of the action to have the correct normalisation.
The constraints 
\beq
\d_0 \tilde{X} = R^2 \d_1 X  \qquad \d_0 X = R^{-2} \d_1 \tilde{X} 
\eeq
 then follow after integrating the equations of motion\footnote{We fix
   the arbitrary function of $\tau$ introduced by integration by
   observing that (\ref{eqTseyaction}) has $\dl{X} = f(\tau)$ gauge
   invariance.} and the string wave equation is given by combining the
 constraint equations.\par
Returning to the general case, the PST procedure yields an action
\bea\label{Daction}
S= \frac{1}{2} \int d^2\sigma\left[ -\M_{\a\b} \partial_1 X^\a \partial_1 X^\b + \L_{\a\b} \partial_1 X^\a \partial_0 X^\b + \K_{\a\b} \partial_0 X^\a \partial_0 X^\b\right]\,   ,
\eea
where $X^\a = (\X^A, Y^a)=(X^i, \tilde{X}_j, Y^a)$ and\footnote{In our notation we reserve the Greek characters $\mu$ and $\nu$ to denote worldsheet indices.}
\beq
\M=
\left(
\begin{array}{cc}
\H  &   0 \\
 0 &    g 
\end{array}
\right),
\L=
\left(
\begin{array}{cc}
L &   0 \\
 0 &    0
\end{array}
\right),
\K=
\left(
\begin{array}{cc}
0  &   0 \\
 0 &    g
\end{array}
\right).
\eeq
For the fibre coordinates we have the equation of motion 
\bea
\d_1 \left( \H \d_1 \X \right) = L \d_1\d_0 \X,
\eea
which integrates to give the constraint (\ref{eqConstraint}).  
This form of the action will be our starting point for calculating the beta-functional of the theory.

\section{The background-field expansion}

To perturbatively study ultra-violet divergences in the doubled
formalism we expand  quantum fluctuations around a classical
background. We make a choice of coordinates that leaves the general
coordinate invariance of the action manifest in the perturbative
expansion\cite{BFM,BCZ}. 

First one writes the fields $X^\a$ as the sum of a classical piece
$X^\a_{cl}$ and a quantum fluctuation $\pi^\a$, however, the expansion
in this fluctuation would not be general covariant. Instead one takes
the geodesic from $X^\a_{cl}$ to  $X^\a_{cl}+\pi^\a$ and finds its
tangent vector at  $X^\a_{cl}$ with length equal to that of the
geodesic. We call this tangent vector $\xi^\a$ and it is contravariant,
so the coefficients in at expansion in terms of $\xi^\a$ are
tensors. In general, terms linear in $\xi^\a$ are proportional to
the equations of motion of the classical background and so
vanish. The quadratic terms will give
rise to the one loop corrections and so are the relevant terms for
calculating the one-loop beta-function. 

We will use the algorithmic method of calculating the background-field expansion developed in \cite{Mukhi}. To obtain the $n$th order background-field expanded action we simply act on the Lagrangian with the operator
\beq
\int d^2\c\xi^\a(\uc )D^\c_\a
\eeq
$n$ times and divide by $n!$ ($D^\c_\a$ is the covariant functional derivative with respect to $X^\a(\uc)$). The action of this operator can be summarised as
\bea
\int d^2\c\,\xi^\a(\uc)D^\c_\a \xi^\b(\uc^\p)&=&0\, ,\\
\int d^2\c \,\xi^\a(\uc)D^\c_\a \d_\m X^\b(\uc^\p)&=&D_\m\xi^\b(\uc^\p)\, ,\\
\int d^2\c \,\xi^\a(\uc)D^\c_\a D_\m\xi^\b(\uc^\p)&=&R^\b_{\phantom{\b}\a\g\dl}\d_\m X^\dl \xi^\a \xi^\g(\uc^\p)\, ,\\
\int d^2\c \,\xi^\a(\uc)D^\c_\a T_{\a_1\a_2\ldots \a_n}(X(\uc^\p))&=&D_\b T_{\a_1\a_2\ldots \a_n}\xi^\b(\uc^\p)\, ,\label{DT}
\eea
where $R^\b_{\phantom{\b}\a\g\dl}$ is the target space Riemann tensor and $T_{\a_1\a_2\ldots \a_n}$ is a rank $n$ tensor and these are understood to be evaluated at the classical value $X_{cl}$. The form of (\ref{DT}) is particularly relevant, leading to simplification when dealing with the metric.

Expanding the first term in (\ref{Daction}) is exactly the same as the
standard sigma model calculation (albeit without world sheet
covariance), at first order we have\footnote{From now on $X$ refers to the classical field $X_{cl}$.}
\beq
-\M_{\a\b}\d_1X^{\a}D_1\xi^\b
\eeq
and at second order
\beq
-\frac{1}{2}\left(\M_{\a\b}D_1\xi^\a D_1\xi^\b+R_{\g\a\b\dl}\xi^\a\xi^\b\d_1X^{\g}\d_1X^\dl\right),
\eeq
where $R_{\g\a\b\dl}$ is the Riemann tensor constructed from the metric $\M$. The expansion of the $\L$ term in the action is more complex giving
\beq
\frac{1}{2}
\left(\L_{\a\b}\d_0X^{\a}D_1\xi^\b+\L_{\a\b}D_0\xi^{\a}\d_1X^\b+D_{\a}\L_{\g\b}\xi^\a\d_0 X^{\g}\d_1X^\b\right)
\eeq
at first order and
\bea
\frac{1}{2}\left(\L_{\a\b}D_0\xi^{\a}D_1\xi^\b+\frac{1}{2}\left( D_\a D_\b\L_{\g\dl} + L_{\g\s}R^{\s}_{\ \a\b\dl} +   L_{\dl\s}R^{\s}_{\ \a\b\g} \right)\xi^\a\xi^\b\d_0X^{\g}\d_1X^\dl \qquad\right.\non\\
+D_{\g}\L_{\a\b}\xi^\g\left(\d_0X^{\a}D_1\xi^\b+D_0\xi^{\a}\d_1X^\b\right)\biggr)
\eea
at second-order. This is the general expansion for any second rank tensor $\L$  so
the $\K$ term in (\ref{Daction}) may be expanded in a similar way. 

The first-order terms in $\xi$ vanish as they should (using the
equations of motion of $X_{cl}$) leaving the second-order Lagrangian which is given by
\bea\label{fulag}
2{\mathcal L}_{(2)}&=&-\M_{\a\b}D_1\xi^\a D_1\xi^\b+\L_{\a\b}D_0\xi^\a D_1\xi^\b+\K_{\a\b}D_0\xi^\a D_0\xi^\b\non\\
&&-R_{\g\a\b\dl}\xi^\a\xi^\b\d_1X^\g\d_1X^\dl+\L_{\a\b;\g}\xi^\g(D_0\xi^\a\d_1X^\b+\d_0X^\a D_1\xi^\b)\non\\
&& +\frac{1}{2}D_\a D_\b\L_{\g\dl}\xi^\a\xi^\b\d_0X^{\g}\d_1X^\dl
+\frac{1}{2}\left( L_{\g\s}R^{\s}_{\ \a\b\dl} + L_{\dl\s}R^{\s}_{\ \a\b\g}\right)\xi^\a\xi^\b\d_0X^\g\d_1X^\dl\non\\
&&+2\K_{\a\b;\g}\xi^\g D_0\xi^\a\d_0X^\b\non\\
&&  +\frac{1}{2}D_\a D_\b\K_{\g\dl}\xi^\a\xi^\b\d_0X^{\g}\d_0X^\dl
+\K_{\g\s}R^{\s}_{\ \a\b\dl}\xi^\a\xi^\b\d_0X^\g\d_0X^\dl \, .
\eea

\section{Simplification strategy}

Now we have the background-field expanded action we may 
simplify it using the equations of motion of $X^{\alpha}$ (recall
$X^{\alpha}$ is the classical field configuration which we are expanding
around and so obeys its equation of motion). Then we use
Wick contractions on the quantum field $\xi$. The equation of motion is 
\beq
D_{1}(\M_{\a\b}\d_1 X^{\b})=\L_{\a\b}\d_1\d_0X^\b+\Dh_{0}(\K_{\a\b}\d_0X^{\b})\, ,
\eeq
where $\Dh_{0}$ is a covariant derivative constructed only from the
base metric $g$ $(\Dh_{0}\K=0$ and $\hat{\Gamma}$ will similarly refer
to the connection constructed from $g$). This equation is required to show vanishing of the first-order
action and we now use it to remove all $\L$ terms (except the $\L$
fluctuation `kinetic' term) from the action. This means (using
integration by parts) we can substitute
\bea
&&\L_{\a\b;\gamma}\xi^{\gamma}\left(\d_0\xi^{\a}\d_1X^{\b}+\d_1\xi^{\a}\d_0X^{\b}\right)+\L_{\a\b}\G{\a}{\g}{\dl}\xi^{\gamma}\d_1X^{\dl}\d_0\xi^{\b}+\L_{\a\b}\G{\a}{\g}{\dl}\xi^{\gamma}\d_0X^{\dl}\d_1\xi^{\b}\non\\
&=&\frac{1}{2}\L_{\b\dl}\d_{\s}\G{\dl}{\g}{\a}\xi^{\g}\xi^{\a}(\d_1X^{\b}\d_0X^{\s}+\d_0X^{\b}\d_1X^{\s})\non\\
&&-2\M_{\dl\b}\G{\dl}{\g}{\a}\xi^{\g}\d_1\xi^{\a}\d_1X^{\b}-\d_{\s}(\M_{\dl\b}\G{\dl}{\g}{\a})\xi^{\g}\xi^{\a}\d_1X^{\b}\d_1X^{\s}\non\\
&&+2\K_{\dl\b}\G{\dl}{\g}{\a}\xi^{\g}\d_0\xi^{\a}\d_0X^{\b}+\d_{\s}(\K_{\dl\b}\G{\dl}{\g}{\a}\xi^{\g})\xi^{\a}\d_0X^{\b}\d_0X^{\s}\non\\
&&+\G{\dl}{\g}{\a}\xi^{\g}\xi^{\a}\left(\M_{\dl\b}\G{\b}{\s}{\tau}\d_1X^{\s}\d_1X^{\tau}-\K_{\dl\b}\hat{\Gamma}^{\b}_{\  \s\tau}\d_0X^{\s}\d_0X^{\tau}\right)\, .
\eea
This leads to a dramatically simplified Lagrangian given by
\bea\label{Lag}
2{\mathcal L}_{(2)}&=&-\M_{\a\b}\d_1\xi^\a \d_1\xi^\b+\L_{\a\b}\d_0\xi^\a \d_1\xi^\b+\K_{\a\b}\d_0\xi^\a\d_0\xi^\b\non\\
&&-2\d_\a\M_{\g\b}\d_1X^\g\xi^\a\d_1\xi^\b-\frac{1}{2}\d_\a\d_\b\M_{\g\dl}\d_1X^\g\d_1X^\dl\xi^\a\xi^\b\non\\
&&+2\d_\a g_{\b\g}\xi^\a\d_0\xi^\b\d_0X^\g+\frac{1}{2}\d_\a\d_\b
\K_{\g\dl}\xi^\a\xi^\b\d_0X^\g\d_0X^\dl  \, \, .
\eea

Note that we have chosen to proceed by expanding covariant derivatives
and simplifying using the equations of motion, rather than leaving
things expressed in terms of covariant derivatives. 
We now proceed to introduce vielbeins so that we can work in the
chiral frame where we know how to find the fluctuation propagators. 

Once vielbeins are introduced there are
of course terms with derivatives acting on the vielbeins. Normally such terms are accounted for by exchanging the usual connection for the spin connection.  The pull back of the spin connection to the world sheet transforms as a gauge field.
 There is then a general argument that this gauge field, 
which is minimally coupled, cannot contribute to the Weyl anomaly. 
We have a modified action where the gauge connection is no 
longer be minimally coupled and there is no such argument 
(indeed we find contributions from the `gauge' terms). 

Introducing the vielbein has the effect of moving all 
indices on $\xi's$ in the second-order action to the chiral 
frame at the expense of also introducing the following terms:
\bea\label{VLag}
2{\mathcal L}_{\V}&=&-2\M_{\ab\b}\d_1\Vt{\b}{\bb}\xi^{\bb}\d_1\xi^{\ab}-\M_{\a\b}\d_1\Vt{\b}{\bb}\d_1\Vt{\a}{\ab}\xi^{\bb}\xi^{\ab}\non\\
&&+\L_{\ab\b}\d_1\Vt{\b}{\bb}\xi^{\bb}\d_0\xi^{\ab}+\L_{\ab\b}\d_0\Vt{\b}{\bb}\xi^{\bb}\d_1\xi^{\ab}+\L_{\a\b}\d_1\Vt{\b}{\bb}\d_0\Vt{\a}{\ab}\xi^{\bb}\xi^{\ab}\non\\
&&+2\K_{\ab\b}\d_0\Vt{\b}{\bb}\xi^{\bb}\d_0\xi^{\ab}+\K_{\a\b}\d_0\Vt{\b}{\bb}\d_0\Vt{\a}{\ab}\xi^{\bb}\xi^{\ab}\non\\
&&-2\d_{\ab}\M_{\g\b}\d_1X^\g\d_1\Vt{\b}{\bb}\xi^{\ab}\xi^{\bb}+2\d_{\ab}
\K_{\g\b}\d_0X^\g\d_0\Vt{\b}{\bb}\xi^{\ab}\xi^{\bb} \, \, .
\eea

\section{Wick contraction}

In this chiral frame $\L$ and $\H$ are diagonal and one can calculate the propagators for the fluctuations from the `kinetic terms' in the Lagrangian.  On the fibre these `kinetic terms' are FJ style Lagrangians of the form (\ref{FJL}) 
for $n$ chiral and $n$ anti-chiral Bosons in flat space.  The propagators for such chiral Lagrangians have previously been considered by 
Tseytlin\cite{Tseytlin1}.  The sum of a chiral and anti-chiral propagator is proportional to a standard Boson propagator $\Delta_0$. The difference of chiral and anti-chiral propagators gives a phase $\theta$. Full details of this are given in the appendix. The general result for our action is that
\beq
\label{prop1}
\langle \xi^{\Ab} \xi^{\Bb} \rangle = \Delta_0\H^{\Ab\Bb}  +\theta L^{\Ab\Bb}.
\eeq

$\Delta_0$ contains UV divergence that needs regularisation and
renormalisation. The coefficients of $\Delta_0$ will thus contribute to the Weyl anomaly and in turn, to the beta-functionals. 

$\theta$ does not contain any divergence and does not therefore
contribute to the Weyl anomaly. Instead  $\theta$ parameterises any
Lorentz anomaly. 
One would demand such anomaly vanishes by setting $\frac{\d  S_{eff}}{\d \theta} = 0$. 
This would place additional constraints on the background fields
beyond that of the  beta-functionals vanishing. However, when
the dust settles all occurrences of $\theta$  in the effective action cancel out leaving no Lorentz anomaly. This is as expected since we have an equal number of Bosons of each
chirality.

Given (\ref{prop1}) we can deduce the form of more complicated Wick
contractions which are quartic in $\xi$ and contain derivatives of fluctuations\footnote{See appendix.}. These contractions arise in the $O(S^2_{eff})$ term in the exponential of the effective action and must be included since we need to count all logarithmic divergences (see for example\cite{BCZ}). 

When evaluating all contributing terms it will be useful to 
distinguish  `base terms' which contain only the base metric $g$ and 
its derivatives and `vielbein terms' which come from (\ref{VLag}) 
and contain derivatives of the vielbein prior to any integrations by parts.

\subsection{Single contraction terms}

The terms with a single $\left<\xi\xi\right>$ contraction are
\beq
\frac{1}{2}\d_a\d_b g_{bg}\left<\xi^a\xi^b\right>\d_\mu Y^g\d^\mu Y^d=\frac{1}{2}\d^a\d_a g_{gd}\d_\mu Y^g\d^\mu Y^d\Delta_0 
\eeq
from the base,
\beq
-\frac{1}{2}\d_a\d_b\H_{GD} \d_1\X^G\d_1\X^D\left<\xi^a\xi^b\right>=-\frac{1}{2}\d_a\d^a\H_{GD} \d_1\X^G\d_1\X^D\Delta_0 
\eeq
on the fibre and
\bea
-\H_{AB}\d_1\Vt{B}{\Bb}\d_1\Vt{A}{\Ab}\left<\xi^{\Bb}\xi^{\Ab}\right>&=&\left(-\d_1\Vt{A}{\Ab}\d_1\Vt{\Ab}{A}+\frac{1}{2}\d_1\H_{AB}\d_1\H^{AB}\right)\Delta_0,\\
L_{AB}\d_1\Vt{B}{\Bb}\d_0\Vt{A}{\Ab}\left<\xi^{\Bb}\xi^{\Ab}\right>&=&\left(\d_1\Vt{\Ab}{A}\d_0\Vt{A}{\Bb}\delta^{\Bb\Cb}L_{\Cb\Ab}\right)\Delta_0,\\
-2\d_{\ar}\H_{BG}\d_1\X^G\d_1\Vt{B}{\Bb}\left<\xi^{\ar}\xi^{\Bb}\right>&=&0,\\
g_{ab}\d^\mu\Vt{b}{\br}\d_\mu\Vt{a}{\ar}\left<\xi^{\br}\xi^{\ar}\right>&=&\left(\d^\mu\Vt{a}{\ar}\d_\mu\Vt{\ar}{a}-\frac{1}{2}\d^\mu g_{ab}\d_\mu g^{ab}\right)\Delta_0,\\
2\d_{\ar}g_{bg}\d^\mu Y^g\d_\mu \Vt{b}{\br}\left<\xi^{\ar}\xi^{\br}\right>&=&\left(-2\d_ag_{gb}\d_d\Vt{b}{\br}\dl^{\ar\br}\V_{\ar}^{\ a}\right)\Delta_0
\eea
from the vielbeins.

\subsection{Double contraction terms}

These occur when expanding the exponential of the effective action to
second order. 
Although there seem myriad possible terms that could contribute,
especially from vielbein terms, many vanish trivially. This because new
divergent diagrams must still be one-loop 
in fluctuations so contain one `loop' of indices: the block diagonal 
form of the metrics and vielbeins ensure the terms mainly 
separate into base and fibre terms, with a few `cross-terms'. 
We use the propagator contractions described in Appendix A and 
note that these terms are a factor of a half down due the exponential,
and a further factor of a half down due to the two sitting on the left-hand side of (\ref{Lag}).

On the base we get
\bea
&&-\d_{\ar}g_{g \br}\d^\mu Y^g\d_{\bar{c}}g_{ d \er}\d_\mu Y^{d}\left<\xi^{\ar}\d_1\xi^{\br}\xi^{\bar{c}}\d_1\xi^{\er}\right>\non\\&&\qquad\qquad\qquad\qquad\qquad=-\frac{1}{2}\left(\d_ag_{gb}g^{bc}\d^ag_{cd}-\d^ag_{bg}\d^bg_{ad}\right)\d^\mu Y^g\d_\mu Y^d\Delta_0
\eea
and on the fibre
\bea
&&\d_{\ar}\H_{G \Bb}\d_1\X^G\d_{\bar{c}}\H_{ D \Eb}\d_1\X^{D}\left<\xi^{\ar}\d_1\xi^{\Bb}\xi^{\bar{c}}\d_1\xi^{\Eb}\right>\non\\&&\qquad\qquad\qquad\qquad\qquad\qquad\qquad=\frac{1}{2}\d_a\H_{GB}\H^{BC}\d^a\H_{CD}\d_1\X^G\d_1\X^D\Delta_0.
\eea

Purely from the the vielbein piece of the Lagrangian (\ref{VLag}) we have 
\bea
\frac{1}{4}L_{\Ab B}\d_1\Vt{B}{\Bb}L_{\Cb
  D}\d_1\Vt{D}{\Db}\left<\xi^{\Ab}\d_0\xi^{\Bb}\xi^{\Cb}\d_0\xi^{\Db}\right>&=&\left(-\d_1\Vt{A}{\Ab}\d_1\Vt{\Ab}{A}+\frac{1}{8}\d_1\H_{AB}\d_1\H^{AB}\right)\Delta_0, \nonumber \\
\frac{1}{4}L_{\Ab B}\d_0\Vt{B}{\Bb}L_{\Cb
  D}\d_0\Vt{D}{\Db}\left<\xi^{\Ab}\d_1\xi^{\Bb}\xi^{\Cb}\d_1\xi^{\Db}\right>&=&\left(-\frac{1}{8}\d_0\H_{AB}\d_0\H^{AB}\right)\Delta_0,\nonumber \\
\frac{1}{2}L_{\Ab B}\d_0\Vt{B}{\Bb}L_{\Cb
  D}\d_1\Vt{D}{\Db}\left<\xi^{\Ab}\d_1\xi^{\Bb}\xi^{\Cb}\d_0\xi^{\Db}\right>&=&\left(-\d_1\Vt{\Ab}{A}\d_0\Vt{A}{\Bb}\hat{\delta}^{\Bb}_{\Ab}\right)\Delta_0, \nonumber \\
-\H_{\Ab B}\d_1\Vt{B}{\Bb}L_{\Cb
  D}\d_1\Vt{D}{\Db}\left<\xi^{\Ab}\d_1\xi^{\Bb}\xi^{\Cb}\d_0\xi^{\Db}\right>&=&\left(2\d_1\Vt{A}{\Ab}\d_1\Vt{\Ab}{A}-\frac{1}{2}\d_1\H_{AB}\d_1\H^{AB}\right)\Delta_0, \nonumber \\
-\H_{\Ab B}\d_1\Vt{B}{\Bb}L_{\Cb
  D}\d_0\Vt{D}{\Db}\left<\xi^{\Ab}\d_1\xi^{\Bb}\xi^{\Cb}\d_1\xi^{\Db}\right>&=&0,\nonumber \\
\H_{\Ab B}\d_1\Vt{B}{\Bb}\H_{\Cb
  D}\d_1\Vt{D}{\Db}\left<\xi^{\Ab}\d_1\xi^{\Bb}\xi^{\Cb}\d_1\xi^{\Db}\right>&=& 0, \nonumber \\
\sum_{\mu=0,1} g_{\ar b}\d^\mu\Vt{b}{\br}g_{\bar{c} d}\d^\mu\Vt{d}{\dr}\left<\xi^{\ar}\d_\mu\xi^{\br}\xi^{\bar{c}}\d_\mu\xi^{\dr}\right>&=& \left(\frac{1}{4}\d^\mu g_{ab}\d_\mu g^{ab}-\d^\mu\Vt{a}{\ar}\d_\mu\Vt{\ar}{a}\right)\Delta_0,
\eea
with one cross-term
\bea
&&\sum_{\mu=0,1} 2\d_{a}g_{gb}\Vt{b}{\br}\Vt{a}{\ar}\d^\mu Y^g g_{\bar{c} e}\d^\mu\Vt{e}{\dr}\left<\xi^{\ar}\d_\mu\xi^{\br}\xi^{\bar{c}}\d_\mu\xi^{\dr}\right>\non\\&&\qquad\qquad\qquad\qquad\qquad=\left(-2\d_ag_{gb}\d_d\Vt{b}{\br}\dl^{\ar\br}\V_{\ar}^{\ a}+\d_ag_{gb}\d_dg^{ba}\right)\d^\mu Y^g\d_\mu Y^d \Delta_0.
\eea

\subsection{The Weyl divergence}

The total Weyl divergence will be given by the coefficient of $\Delta_0$ which we denote by $W$ so that
\beq
S_{Weyl}=\frac{1}{2}\int d^2 \c \bigl[ -W_{GD} \d_1 \X^G\d_1 \X^D + W_{gd}\d_\mu Y^g \d^\mu Y^d \bigr]\Delta_0 \, .
\eeq
 On the base the divergence $W_{gd}$ is given by 
\bea\label{Weylb1}
W_{gd}&=&\frac{1}{2}\d^a\d_a g_{gd}-\frac{1}{4}\d_g g_{ab}\d_d g^{ab}-\frac{1}{2}\d_ag_{gb}g^{bc}\d^ag_{cd}\non\\&&+\frac{1}{2}\d^ag_{bg}\d^bg_{ad}+\d_ag_{gb}\d_dg^{ba}\non\\
&&-\frac{1}{8}\d_g\H_{AB}\d_d\H^{AB}.
\eea
The divergence on the fibre is 
\beq\label{fibdiv}
W_{GD}=\frac{1}{2}\d^2\H_{GD}-\frac{1}{2}\left((\d_a\H)\H^{-1}(\d^a\H)\right)_{GD}.
\eeq
The divergence on the base, (\ref{Weylb1}), can be rewritten as
\bea\label{Weylb2}
W_{gd}&=&g^{ab}g_{gs}\left(\d_b \G{s}{a}{d}+\G{s}{b}{t}\G{t}{a}{d}\right)\non\\
&&-\frac{1}{2}\d_g g_{ab}\d_d g^{ab}+\d_ag_{gb}\d_dg^{ba}\non\\
&&-\frac{1}{8}\d_g\H_{AB}\d_d\H^{AB},
\eea
where we recognise the first two terms as part of the Ricci tensor. We
now, using the base components of the equation of motion for the fields, add zero to the divergence in the form 

\beq
\label{eom0} 
\G{t}{a}{b}g^{ab}\left(\Dh_\mu\left(g_{td}\d^\mu Y^d \right)-\frac{1}{2}\d_t\H_{GD}\d_1 \X^G\d_1\X^D\right)\, . 
\eeq

The base divergence becomes
\bea\label{Weylb3}
W_{gd}&=&-\hat{R}_{gd}-\frac{1}{8}\d_g\H_{AB}\d_d\H^{AB},
\eea
where $\hat{R}_{gd}$ is the ricci tensor constructed from the base metric $g$ alone. The fibre components of the divergence become
\beq
W_{GD}=\frac{1}{2}\d^2\H_{GD}-\frac{1}{2}\left((\d_a\H)\H^{-1}(\d^a\H)\right)_{GD}-\frac{1}{2}\G{t}{a}{b}g^{ab}\d_t\H_{GD}\,  \label{Weylf}
\eeq

\subsection{Relation to the doubled Ricci tensor}

If we calculate the Ricci tensor of the doubled space (the Ricci tensor of $\M$) for comparison, and drop terms proportional to $\H^{AB}\d_d \H_{AB}=0$, it also has block diagonal form with
\bea
R_{GD}&=&-\frac{1}{2}\d^2\H_{GD}+\frac{1}{2}\left((\d_a\H)\H^{-1}(\d^a\H)\right)_{GD}+\frac{1}{2}\G{t}{a}{b}g^{ab}\d_t\H_{GD}\, ,\label{Ricci}\\
R_{gd}&=&\hat{R}_{ab}+\frac{1}{4}\mbox{tr}(\d_g\H\d_d\H^{-1})
\, ,
\eea
for the fibre and base parts respectively. We see that the Weyl divergence are almost equal to minus the Ricci tensor  
except that the term on the base containing the doubled metric $\H$
have an extra factor of $1/2$. 
We note also that the fibre divergence is contracted with
$\d_1\X\d_1\X$, whereas if we considered an ordinary sigma model with
metric $\M$ the fibre piece would be contracted with $\d^\mu\X\d_\mu\X$. 
However, we can use the fibre equations of motion to make $W_{GD}$
contract $\d^\mu\X\d_\mu\X$ at the expense of introducing a factor of $1/2$.
Then, comparing $W$ with $R$, all terms containing the doubled metric 
$\H$ would be a factor of $1/2$ down. 
We will see that writing the doubled metric in terms of the 
standard sigma model fields $h$ and $b$ takes care of these extra factors.

\section{Doubled renormalisation}

One may now proceed directly to regularise and renormalise the divergences coming from $\Delta_0$.  In the standard way one would dimensionally regularise and introduce a mass scale $\mu$  through, say, minimal subtraction and the introduction of counter terms.  We then absorb all scale dependence to define the renormalised couplings
\bea
\left\{\M, \K, \L \right\} \rightarrow \left\{\M^R(\mu), \K^R(\mu),\L^R(\mu)\right\}
\eea
producing the renormalised action
\bea\label{DactionRen}
S^R= \frac{1}{2} \int d^2\sigma\left[ -\M^R(\mu)_{\a\b} \partial_1 X^\a \partial_1 X^\b + \L^R(\mu)_{\a\b} \partial_1 X^\a \partial_0 X^\b + \K^R(\mu)_{\a\b} \partial_0 X^\a \partial_0 X^\b\right]\,   .
\eea
We can calculate the beta-functions from this by differentiating the renormalised couplings with respect to the log of the mass scale giving
\bea
\b^\M_{\a\b} = -\left(\begin{array}{cc}
W_{AB} & 0 \\ 0 & W_{ab}
\end{array}\right)\, , & \b^\K_{\a\b} = -\left(\begin{array}{cc}
0 & 0 \\ 0 & W_{ab} 
\end{array}\right)\, , & \b^\L_{\a\b} = 0 \, .
\eea
 Demanding the vanishing of these beta-functions gives the background field equations. 


\section{Equivalence with standard sigma model}
Instead of working directly with these doubled beta-functions we shall show the equivalence to the standard sigma model by expanding out the Weyl divergence in terms of the non-doubled metric and B-field and eliminate the extra doubled coordinates before renormalisation.  Since this can be cast in a well known form for trivial base metric we will proceed putting $g_{ab}=\dl_{ab}$. Expanding the Weyl divergence $W_{\a\b}$ using (\ref{H}) we obtain
\bea
W_{AB} &=& \frac{1}{2}\left(\begin{array}{cc}\left( r + bh^{-1}r h^{-1}b  - bh^{-1}s  - sh^{-1}b\right)_{ij}  & \left(sh^{-1} -  bh^{-1} r h^{-1}\right)_i{}^j\\ -\left(h^{-1}s-   h^{-1}r bh^{-1}\right)^i{}_j & -\left(h^{-1}rh^{-1}\right)^{ij}\end{array}\right),\\
W_{ab} &=&  \frac{1}{4} t_{ab},
\eea
where we have defined
\bea
r_{ij} &=& \left(\d^2 h -  \d_a h h^{-1} \d^a h - \d_a b h^{-1} \d^a b\right)_{ij},\\
s_{ij} &=& \left(\d^2 b -  \d_a b h^{-1} \d^a h - \d_a h h^{-1} \d^a b\right)_{ij},\\
t_{ab} &=&  \mbox{tr}\left( h^{-1}\d_a h h^{-1}\d_b h - h^{-1}\d_a b h^{-1} \d_b b\right).
\eea
Recall that $X^A = \X^A = \left(X^i, \tilde X_i\right)$. We now wish to eliminate the dual coordinates $\tilde{X}_i$ from the Weyl divergence using the constraint 
\beq
d \X^A = L^{AB}\H_{BC}\ast d \X^C ,
\eeq
which implies that
\bea
 \d_1\tilde{X}_i&=& h_{ij}\d_0X^j + b_{ij}\d_1X^j \,.
\eea
We can observe that the right hand side of the above has a sensible interpretation in terms of the standard sigma model; it is proportional to the canonical momentum.  On using the constraint we find that
\beq
W_{AB}\d_1 \X^A\d_1 \X^B = \frac{1}{2}r_{ij}\d_\mu X^i\d^\mu X^j + \frac{1}{2}\e^{\mu\nu}s_{ij}\d_\mu X^i\d_\nu X^j.
\eeq
Thus, we find that prior to renormalisation, the Weyl divergence part of the effective action is
\bea
\label{eqSeffdoubled}
S_{Weyl} &=&  \frac{1}{2} \int d^2\s\left[ -W_{AB}\d_1 \X^A\d_1 \X^B + W_{ab}\d_\mu Y^a \d^\mu Y^b\right]\Delta_0 \\ 
&=&  \frac{1}{4} \int d^2\s \left[ r_{ij}\d_\mu X^i \d^\mu X^j  +  s_{ij}\e^{\mu\nu} \d_\mu X^i\d_\nu X^j + \frac{t_{ab}}{2}\d_\mu Y^a \d^\mu Y^b\right]\Delta_0.\label{dea}
\eea
Demanding that this divergence vanishes constrains the background fields to obey $r = s = t =0$.\par
We now wish to compare this to the standard sigma model in conformal gauge 
\bea
S=\frac{1}{2}\int d^2\sigma \left[ G_{\a\b}\d_\mu X^\a \d^\mu  X^\b  + \e^{\mu\nu}B_{\a\b}\d_\mu X^\a \d_\nu  X^\b\right] \eea
with metric and B-field
\bea
\label{eqAnsatz}
G_{\a\b} = \left(\begin{array}{cc}
h_{ij}(Y) & 0 \\ 0 & \dl_{ab}
\end{array}\right) , & B_{\a\b} = \left(\begin{array}{cc}
b_{ij}(Y) & 0 \\ 0 & 0
\end{array}\right).
\eea
The beta-functionals for this sigma model are \cite{Callan:1985ia}\footnote{We have set the dilaton to a constant.}
\bea
\b^G_{\a\b} &=& R_{\a\b} -\frac{1}{4}H_{\a\c\dl}H_{\b}^{\phantom{\a}\c\dl}, \\
\b^B_{\a\b} &=& - \frac{1}{2} D^{\c}H_{\c\a\b},
\eea
where $H_{\a\b\c} = \d_\a B_{\b\c} + \d_\a B_{\b\c} +\d_\a B_{\b\c} $.  On substitution of our ansatz for $B$ and $G $ we find that the non-vanishing components are
\bea \label{eqSinglebeta1}
\b^{G}_{ij} &=& -\frac{1}{2}\left( r + \frac{1}{2} tr\left(h^{-1}\d_a
h\right)\d^ah\right)_{ij},\label{bGf} \\
\b^{G}_{ab} &=&  -\frac{1}{2}\left( \frac{t_{ab}}{2} + \d_a tr\left(h^
{-1}\d_{b} h \right)  \right) \label{bGb},\\
\b^{B}_{ij} &=& -\frac{1}{2} \left( s+ \frac{1}{2} tr\left(h^{-1}\d_a
h\right) \d^a b\right)_{ij},\label{bBf}\\
\b^{B}_{ab} &=& \frac{1}{4} tr \left(  h^{-1}\d_a b h^{-1} \d_b h + h^
{-1}\d_a h h^{-1} \d_b b    \right).
\eea
The Weyl divergent part of the effective action which produces these beta-functions after renormalisation is, 
\bea\label{eqSinglebfe}
S_{Weyl} = -\frac{1}{2} \int d^2 \s \bigl[\b^{G}_{ij}\d_\mu X^i \d^\mu X^j &+& \b^{G}_{ab}  \d_\mu Y^a \d^\mu Y^b \non\\
&+& \e^{\mu\nu}\left( \b^{B}_{ij}\d_\mu X^i \d_\nu X^j   + \b^{B}_{ab} \d_\mu Y^a \d_\nu Y^b\right)\bigr]\Delta_0.
\eea
The anti-symmetry of $\e^{\mu\nu}\d_\mu Y^a \d_\nu Y^b$ allows us to
cancel the divergence that gave rise to $\b^{B}_{ab}$.  
The equation of motion for the base coordinate $Y$ is 
\bea
2\d^2 Y^a = \d^a h_{ij} \d_\mu X^i \d^\mu X^j  + \d^a b_{ij}\e^{\mu\nu}\d_\mu X^i \d_\nu  X^j , \eea
so upon multiplying both sides $tr\left(h^{-1}\d_a h\right)$ and integrating by parts we have
\bea
&&{}\,\frac{1}{2}\int d^2\s \,\mbox{tr}\left(h^{-1}\d_a h\right)\left(\d^ah_{ij}\dl^{\mu\nu} + \d^ab_{ij}\e^{\mu\nu}\right)\d_\mu X^i\d_\nu X^j  \\
&=&\int d^2\s\,  \mbox{tr}\left(h^{-1}\d_a h\right)\d^2 Y^a = -\int d^2\s \,  \d_a \mbox{tr}\left(h^{-1}\d_{b} h \right)  \d_\mu Y^a\d^\mu Y^b
\eea
so that (\ref{eqSinglebfe}) reduces to
\bea
\label{eqSinglebfe2}
S_{Weyl} =\frac{1}{4} \int \bigl[ r_{ij}\d_\mu X^i \d^\mu X^j  +
s_{ij}\e^{\mu\nu} \d_\mu X^i\d_\nu X^j + \frac{t_{ab}}{2}\d_\mu Y^a
\d^\mu Y^b\bigr]\Delta_0 \, .
\eea
This agrees with what we found previously from the doubled formalism in (\ref{dea}). Thus after integrating out the dual coordinate the doubled formalism gives {\it exactly} the same divergent terms as the standard string sigma model.

The construction can be straightforwardly extended to include a non-trivial base metric $g(Y)$. In
this case the following additional terms
\bea
&&\frac{1}{2}\G{t}{a}{b}g^{ab}\d_t h_{ij}\, ,\\
&&\hat{R}_{ab}+\frac{1}{4}\mbox{tr}(h^{-1}\d^th)\d_tg_{ab}-\frac{1}{2}
\mbox{tr}(h^{-1}\d^th)\d_ag_{tb}\, ,\\
&&\frac{1}{2}\G{t}{a}{b}g^{ab}\d_t b_{ij}\, ,
\eea

are required to reproduce the usual beta-functionals (\ref{bGf}),
(\ref{bGb}) and (\ref{bBf}) respectively. These terms do indeed follow
from the doubled geometry Weyl divergences (\ref{Weylb3}) and (\ref{Weylf}) after application of the equations of motion.
In fact one can immediately see $\hat{R}_{ab}$, the Ricci tensor of the base
metric $g$ in (\ref{Weylb3}).

\section{Conclusion}

In summary, we have been able to calculate the one-loop Weyl divergence
of the duality symmetric string which upon renormalisation gives rise 
to the beta-functionals for the doubled geometry metric couplings.  
For the fibre coordinates these are the obvious geometric quantity,
the doubled target space Ricci tensor.  
For base coordinates, the terms in the Ricci tensor that contain the
fibre metric $\H$ pick up an extra factor of a half.  
However, when we interpreted the results in terms of the non-doubled
fields these factors are taken care of and indeed we are left 
with exactly the same Weyl divergence as for the standard sigma model.  

In this calculation there are some notable features that we wish to
draw attention to.  First, 
the topological term which is crucial in establishing equivalence of
partition functions with the standard sigma model, 
played no role in this calculation.  
Second, the non-covariant structure of the action (\ref{Daction})
meant that unlike the calculation for the standard string, the `gauge'
terms do make a contribution to the divergence and increase the
computational difficulty. Third, the B-field is incorporated into the doubled metric and the is no anti-symmetric term in the action. Fourth, the chiral nature of the fibre coordinates suggests that one should be concerned about any potential Lorentz anomaly. This anomaly
actually vanished, cancelling between the Bosons of opposite chirality.
Finally, we found that the $\L$ coupling containing the  $O(n,n)$ fibre metric $L$ does not get renormalised.

\subsection{Discussion}

There are a number of assumptions in this work that would be
interesting to explore further.
We assumed that the connection in the fibration was identically zero. 
To include such a connection would add off diagonal elements to the 
doubled-space metric and would also require a suitably generalised 
constraint (with derivatives promoted to covariant derivatives).  
On applying the PST procedure to produce an action akin to
(\ref{Daction}) one finds that terms involving the connection appear
in both the metric $\M$ and the `invariant' $O(d,d)$ metric $\L$.  
Evaluation of the divergence in this case would be more challenging.

As with other treatments of the duality symmetric formalism we had to
specify that the fibre metric depended only on the base
coordinates. It would be nice to relax this assumption. The difficulty
with doing so is that  the chirality constraint would have to be
modified as would its implementation with a suitable generalisation of
the PST action. Another interesting and perhaps more democratic 
generalisation along this line would be the doubling of all coordinates. 
It is remarkable that the background field equations obtained required
no use of the presence of any Killing directions implying that the
doubled formalism is more general than one might have been first led
to believe.

In this paper we have not included the doubled space dilaton, $\Phi$, which is related to the standard dilaton, $\phi$, through 
\bea \label{Dilrelation} \Phi = \phi - \frac{1}{2}\ln \det h \, .\eea  
This is introduced into doubled formalism with the usual Fradkin--Tseytlin action
\beq 
S_{dil} = \frac{1}{8\pi} \int d^2 \sigma \Phi(Y) R^{(2)}.
\eeq
The doubled dilaton is T-dual invariant. However integrating out the dual coordinates from the doubled action (\ref{Daction}) produces a determinant factor which correctly reproduces the transformation of the standard dilaton under T-duality \cite{Hull:2006va}.  Similar invariant dilatons have been used in other treatments of T-duality as review in \cite{Giveon:1994fu}.   

The relation between the two dilatons (\ref{Dilrelation}) tells us that if we wish to set the doubled dilaton $\Phi$ to be a constant it is inconsistent to also set  $\phi$ also to be a constant as we did when showing the equivalence of Weyl divergence.  Instead, one should have $\phi =  \frac{1}{2}\ln \det h$ and consider the full beta-functions for the metric and B-field in the standard sigma model
\bea
\b^G_{\a\b} &=& R_{\a\b} + D_\a D_\b \phi \, ,\\
\b^B_{\a\b} &=& -\frac{1}{2} D^\dl H_{\a\b\dl} +  \frac{1}{2} D^\dl \phi H_{\a\b\dl}\, .
\eea
The addition of these dilaton terms actually produces an exact match to the doubled space beta-functions without further manipulation using the equations of motion as we did in Section 7. 

A further generalisation is to consider a general non-constant dilaton in the doubled theory producing a beta-functional $\b^\Phi$. It is clear that the leading term in $\b^\Phi$ will be proportional to
$26-c$ as is the case for the standard string. 
The $26$ comes from an integration over world sheet metrics which
remains unchanged upon doubling the target space. $c$ is the central charge of the theory which remains the
same after doubling since although we now have $d+2n$ Bosons (d being the dimension of the non-doubled space), $2n$ of
these are chiral and so contribute only a half each to the central charge. To be concrete about the full equivalence of the beta-functions with a general dilaton would require extending our analysis
through to two loops. Indeed, higher loop analysis could still provide
interesting corrections that are not present at leading order.

\section{Acknowledgements}

DSB is supported by EPSRC grant GR/R75373/02 and would like to thank DAMTP and Clare Hall college
Cambridge for continued support. NBC would also like to thank DAMTP for continued support. This work was in part supported by the EC Marie Curie 
Research Training Network, MRTN-CT-2004-512194. DCT is supported by a
STFC studentship. We would like to thank Chris Hull, James Lucietti,
Andrew Low and Malcolm Perry for discussions.

\begin{appendix}
\section{Propagators}

To find the propagators for the fluctuations we look at the kinetic terms for the scalars when we have rotated the Lagrangian (\ref{Lag}) to the chiral frame. The fluctuations have either chiral (+) or anti-chiral  (-) Floreanini-Jackiw style kinetic terms with action
\bea
S_{\pm} = \pm\frac{1}{2}\int d^2\s \phi \d_\mp\d_1 \phi\,.
\eea
This action yields momentum-space loop propagators 
\bea\label{Dpm}
\Delta_\pm(\s,\s')  &=& \pm i\int \frac{d^2p}{(2\pi)^2} \frac{1}{p_1p_\mp}e^{-ip.(\s-\s')}, 
\eea 
and we will normally write $\Delta_\pm$ to indicate the $\s\rightarrow\s'$ limit. Simply by examining the integrals we see
\bea
\Delta_+ + \Delta_-& =& 2\Delta_0, \label{D00}\\
\Delta_+ - \Delta_-& =& 2\t\, ,
\eea
where $\Delta_0(\s-\s') =-i \int \dtp \frac{1}{p^2}e^{-ip.(\s-\s')}$ is the propagator for a non-chiral Boson\footnote{We use $(+,-)$ signature on the worldsheet.} and we take this as the definition of $\theta$. Of course these propagator integrals are divergent and we must regularise and then renormalise to find the beta-function.

The propagators can be calculated in $z$-space after Wick rotation\cite{Tseytlin1} with $z=\s+it=\s+\tau$ and $\d_\s=\d+\bar{\d},\d_\tau=\d-\bar{\d}$. Using a $z\rightarrow0$ regularisation such that $\bar{\d}z^{-1}=\pi\delta^{(2)}(z)$ one finds
\bea
\Delta_{+}(z , z^\p ) &=& -\frac{1}{2\pi}\ln ( z- z^\p) , \\
\Delta_{-}(z , z^\p ) &=& -\frac{1}{2\pi}\ln ( \bar{z}- \bar{z}^\p) ,\\
\Delta_+(z , z^\p ) +  \Delta_-(z , z^\p ) &=& -\frac{1}{2\pi}\ln | z- z^\p|^2 = 2\Delta_0( z , z^\p), \label{D0}\\
\Delta_+(z , z^\p ) - \Delta_-(z , z^\p ) &=& -\frac{1}{2\pi}\ln \frac{ z- z^\p}{\bar{z}- \bar{z}^\p}=-\frac{i}{\pi} \mbox{arg}(z-z')=2\t \label{theta} \, ,\eea
where in (\ref{D0}) we have noted after regularisation we have the same relation as (\ref{D00}) to the standardly normalised two-dimensional scalar propagator in this regularisation scheme. 

Terms in the path integral of the effective action that are proportional to $\Delta_0$ will be those related to a breakdown in Weyl invariance whereas terms proportional to $\t$ will correspond to a breakdown in worldsheet Lorentz invariance\cite{Tseytlin1,Tseytlin2}. Looking at the form of (\ref{D0}) and (\ref{theta}) we can see scaling of $z$ shifts $\Delta_0$ and not $\t$ and rotation by a phase shifts $\t$ and not $\Delta_0$. One can also see that in flat space the propagator between two $X$ coordinates or two $\xt$ coordinates is $\Delta_0$ whereas between an $X$ and an $\xt$ it is $\t$. The beta-function should come from terms proportional to $\Delta_0$ in the exponential of the the action and one can obtain it by regularising and renormalising $\Delta_0$ using dimensional regularisation as is standard. Terms proportional to $\t$ would mean additional constraints on the background to ensure worldsheet Lorentz invariance and would indicate a difference from the ordinary sigma model. We find such terms cancel giving agreement with the standard formulation. We would expect this as we have equal numbers of each chirality of Boson.

Looking again at our general $d$-dimensional doubled action for the fibre coordinates in the chiral frame (indicated by barred indices) we have fluctuation `kinetic terms' given by
\bea
S= \frac{1}{2}\int \left[-\H_{\Ab\Bb} \d_1\xi^{\Ab} \d_1\xi^{\Bb}  + L_{\Ab\Bb} \d_1\xi^{\Ab} \d_0\xi^{\Bb}\right],
\eea
where in this frame the metrics are diagonal: 
\bea
\H = \left(\begin{array}{cc}
\openone & 0 \\ 0 & \openone\end{array}
\right), &  L = \left(\begin{array}{cc}
\openone & 0 \\ 0 & -\openone\end{array}
\right).
\eea
Thus the general propagator for $\xi^\Ab$ is given by 
\bea
\langle \xi^{\Ab}(z) \xi^{\Bb}(z) \rangle &=&\left(\begin{array}{cc}
\openone & 0 \\ 0 & 0\end{array}
\right)\Delta_{+}+ \left(\begin{array}{cc}
0 & 0 \\ 0 & \openone\end{array}
\right) \Delta_{-}\\&=& \frac{1}{2}(\H + L) \Delta_{+} + \frac{1}{2}(\H - L) \Delta_{-}\\
&=&  \frac{1}{2}\H^{\Ab\Bb}(\Delta_+ + \Delta_- ) +  \frac{1}{2}L^{\Ab\Bb}(\Delta_+ - \Delta_- ) \\
&=& \Delta_0\H^{\Ab\Bb}  +\theta L^{\Ab\Bb}\, .
\eea
We can use this result and a Wick contraction procedure, described for the ordinary string in \cite{BCZ},
 to determine the divergent behaviour of higher-order propagator contractions which appear in the expansion of the exponential of our action. 

For example
\bea
&&i\int d^2\s'\left< \xi(\s)^\Ab \d_1 \xi(\s)^\Bb \xi(\s')^\Cb \d_0 \xi(\s')^\Db \right>\non\\
&=&i\int d^2\s'\left( \frac{1}{2}(\H + L)^{\Ab\Cb} \Delta_{+}(p) + \frac{1}{2}(\H - L)^{\Ab\Cb}  \Delta_{-}(p)\right)e^{ip.(\s-\s')}\non\\
&&\times\left( \frac{1}{2}(\H +L)^{\Bb\Db} \Delta_{+}(q) + \frac{1}{2}(\H - L)^{\Bb\Db}  \Delta_{-}(q)\right)q_1q_0e^{iq.(\s-\s')}\non\\
&&+i\int d^2\s'\left( \frac{1}{2}(\H + L)^{\Ab\Db} \Delta_{+}(p) + \frac{1}{2}(\H - L)^{\Ab\Db}  \Delta_{-}(p)\right)p_1e^{ip.(\s-\s')}\non\\
&&\times\left( \frac{1}{2}(\H +L)^{\Bb\Cb} \Delta_{+}(q) + \frac{1}{2}(\H - L)^{\Bb\Cb}  \Delta_{-}(q)\right)q_0e^{iq.(\s-\s')}.
\eea
The $\s'$ integral gives $(2\pi)^2\dl(p+q)$ and putting in the forms of the integrals in $\Delta_\pm$ from (\ref{Dpm}) gives 
\bea
\nonumber
&&i \int d^2\s'\left< \xi(\s)^\Ab \d_1 \xi(\s)^\Bb \xi(\s')^\Cb \d_0 \xi(\s')^\Db \right>\\\nonumber
&&= \frac{i}{8} \int \dtp  \left( (\H - L) \left( \frac{1}{p_1} -  \frac{1}{p_+}\right)- (\H + L)\left(\frac{1}{p_1}  + \frac{1}{p_-}\right) \right)^{\Ab\Cb}\\
&&\qquad\qquad\qquad\qquad\qquad\qquad\times   \left( \frac{(\H - L)}{p_+}-\frac{(\H + L)}{p_-}\right)^{\Bb\Db}-  (\Ab \leftrightarrow \Bb)\\
&&\simeq \frac{1}{8}\left(\begin{array} {c}
-\left(\H\H -L\H  - \H L+LL \right) \Delta_- \\
+\left(\H\H +L\H -\H L -LL \right) (-\Delta_- + \Delta_0)\\
+\left(\H\H -L\H +\H L -LL \right) (\Delta_+ - \Delta_0)\\
+ \left(\H^{\Ab\Cb}\H^{\Bb\Db} +L\H +\H L+LL \right) \Delta_+ 
\end{array}\right)  -  (\Ab \leftrightarrow \Bb)\\\nonumber \\  
&&= -\Delta_0\left(\H^{\Ab\Cb}L^{\Bb\Db} +L^{\Ab\Cb}\H^{\Bb\Db} \right) - \t L^{\Ab\Cb}L^{\Bb\Db}-  (\Ab \leftrightarrow \Bb), 
\eea
where $\simeq$ indicates equality up to convergent terms which are irrelevant for our purpose.  

A similar procedure can be used to calculate the two-propagator contractions with any combination of worldsheet derivatives. One can also allow indices on the base; when all four indices are on the base the calculations are as for the standard sigma model (see for example \cite{BCZ}) and since there is no base--fibre propagator the only other allowed possibility  is to have two indices on the base and two on the fibre. The results can be compactly summarised in terms of the total space metric $\M$ and  $\L$ as\footnote{We will simplify notation by using $\left<\xi^\Ab \d_1 \xi^\Bb \xi^\Cb \d_0 \xi^\Db \right>=i\int d^2\s'\left< \xi(\s)^\Ab \d_1 \xi(\s)^\Bb \xi(\s')^\Cb \d_0 \xi(\s')^\Db \right>$ it will always appear in the expansion of the exponential of the effective action in this form.}
\bea
 \langle \xi^\ab \d_1 \xi^\bb \d_1 \xi^\cb \xi^\db \rangle &=& \Delta_0\left(\M \M -\L\L\right)^{(\ab\cb\bb\db - \ab\db\bb\cb)},\\
\langle {\xi^\ab \d_1 \xi^\bb \d_0 \xi^\cb \xi^\dlb} \rangle &=& -\Delta_0\left(\M\L+\L\M\right)^{(\leftrightarrow)} -  \t \L\L^{(\leftrightarrow)},   \\
 \langle {\xi^\ab \d_0 \xi^\bb \d_0 \xi^\cb \xi^\dlb}\rangle &=& -\Delta_0\left(\M\M+3\L\L\right)^{(\leftrightarrow)} -  \t \left( \L\M + \M\L\right)^{(\leftrightarrow)},   
\eea
where $\M \M^{(\ab\cb\bb\db - \ab\db\bb\cb)} $ represents $ \M^{\ab\cb} \M^{\bb\db}  - \M^{\ab\db} \M^{\bb\cb}$ and  $(\leftrightarrow)$ understood in the same way. 
\end{appendix}

\end{document}